\documentclass[journal,twocolumn]{IEEEtran}
\usepackage{amsmath}
\usepackage{amsthm}
\usepackage{multirow}
\usepackage{color,cite}
\usepackage{subfigure}
\usepackage{mathrsfs}
\usepackage{amssymb}
\usepackage{bm}
\usepackage{graphicx}
\usepackage{amsfonts}
\usepackage{algorithm}
\usepackage{algorithmic}
\usepackage{makecell}
\usepackage{etoolbox}

\usepackage[table]{xcolor} 
\usepackage{enumitem}
\usepackage{url}

\allowdisplaybreaks[4]

\input epsf

\begin{document}

\title{{
High-Efficiency Urban 3D Radio Map Estimation Based on Sparse Measurements}}
\author{Xinwei Chen, Xiaofeng Zhong,~\IEEEmembership{Member,~IEEE},  Zijian Zhang,~\IEEEmembership{Graduate Student Member,~IEEE}, \\Linglong Dai,~\IEEEmembership{Fellow,~IEEE}, and  Shidong Zhou,~\IEEEmembership{Member,~IEEE}
 \vspace{-10mm}

\thanks{ X. Chen, Z. Zhang, X. Zhong, L. Dai, and S. Zhou are with Department of Electronic Engineering, Tsinghua University, Beijing 100084, China. (E-mails: \{cxw22, zhangzj20\}@mails.tsinghua.edu.cn, 
\{zhongxf, daill, zhousd\}@mail.tsinghua.edu.cn).}
\vspace{-2mm}
 }

\twocolumn

\maketitle

\begin{abstract}
Recent widespread applications for unmanned aerial vehicles (UAVs) -- from infrastructure inspection to urban logistics -- have prompted an urgent need for high-accuracy three-dimensional (3D) radio maps. However, existing methods designed for two-dimensional radio maps face challenges of high measurement costs and limited data availability when extended to 3D scenarios. To tackle these challenges, we first build a real-world large-scale 3D radio map dataset, covering over 4.2 million $\text{{m}}^3$ and over 4 thousand data points in complex urban environments. We propose a Gaussian Process Regression-based scheme for 3D radio map estimation, allowing us to realize more accurate map recovery with a lower RMSE than state-of-the-art schemes by over 2.5 dB. To further enhance data efficiency, we propose two methods for training point selection, including an offline clustering-based method and an online maximum a posterior (MAP)-based method. Extensive experiments demonstrate that the proposed scheme not only achieves full-map recovery with only 2\% of UAV measurements, but also sheds light on future studies on 3D radio maps.

\end{abstract}


\IEEEpeerreviewmaketitle

\vspace{-3mm}
\section{Introduction}
\vspace{-1mm}

Driven by the burgeoning demand for smart city solutions, unmanned aerial vehicle (UAV) technology has become an integral component. Due to their inherent flexibility, cost-effectiveness, and eco-friendliness, UAVs have demonstrated remarkable growth in applications such as inspection, disaster relief, and logistics. In the logistics field, leading companies such as Meituan and SF Express have widely adopted UAVs, thereby significantly boosting delivery capacity and efficiency across diverse terrains and scenarios.

However, the deployment of UAVs in urban areas encounters considerable challenges in maintaining robust communication during missions, a critical factor for effective flight trajectory planning \cite{deng2023real}. First, navigating low-altitude urban areas is complex due to the impact of dense buildings on communication signal strength and spectrum resources, which vary significantly in three-dimensional (3D) spaces. Second, the high cost of data measurements and the restricted flight accessibility of UAVs result in limited data availability\cite{ivanov2022uav}. Third, there is an urgent need for efficient updates to adapt to the dynamic urban development. Consequently, recovering comprehensive 3D radio maps from sparse UAV measurements becomes indispensable.

Existing works have shown that altitude significantly impacts signal strength and throughput of transceivers, with differences at various heights reaching up to a factor of two\cite{horsmanheimo20225g}. Current open-source radio map datasets are usually two-dimensional, which is inadequate for UAV trajectory planning\cite{0gtx-6v30-22, wmks-4475-21}. Most existing methods for radio map estimation are data-driven, which are high-cost and difficult to achieve comprehensive coverage, especially in 3D scenarios \cite{shrestha2022deep, sato2021space}. Model-driven approaches often rely on oversimplified models, leading to high computational complexity and model errors of about 13.5 dB \cite{thrane2019comparison}. Due to their inherent limitations in capturing the complex signal variability of real-world scenarios, these approaches are typically hard to generalize in new environments or represent signal dead zones \cite{shrestha2022spectrum}.

\begin{figure}[!t]

	\center
	\includegraphics[width=1\linewidth, keepaspectratio]{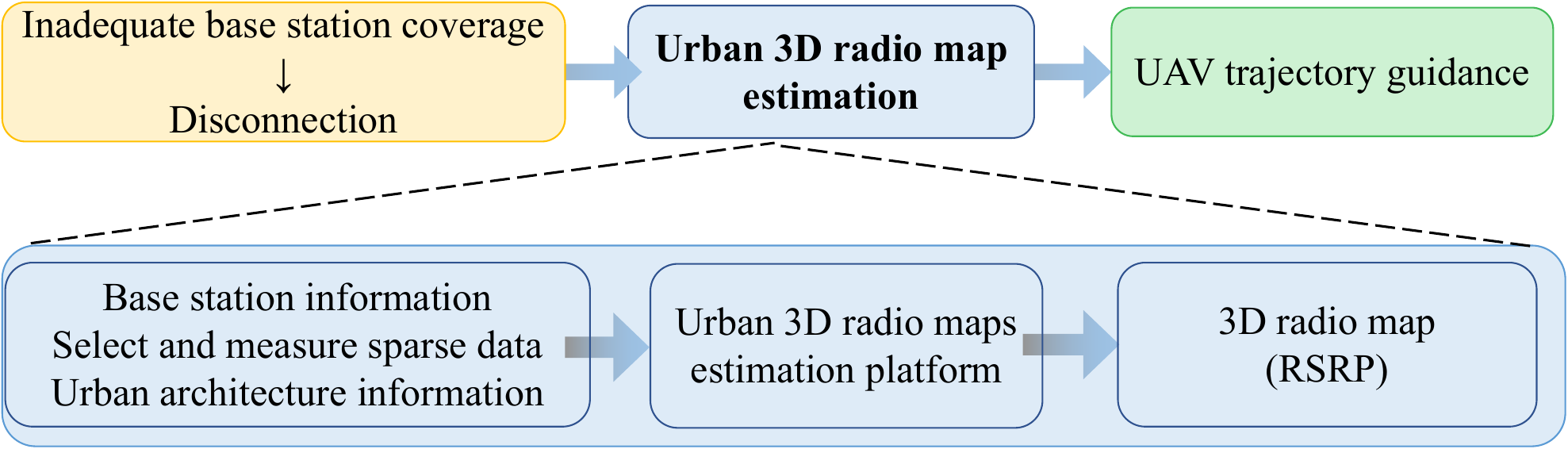}	
     \vspace{-5mm}
        \caption{Motivation and overview of urban 3D radio map estimation procedure.}
	\label{chart}

	\vspace{-6mm}
\end{figure}

To achieve high-efficiency 3D radio map recovery, this paper develops an urban 3D radio maps estimation scheme, which utilizes the sparse data collected by UAVs, as illustrated in Fig. \ref{chart}. We summarize our contributions as follows:

\begin{itemize}[leftmargin=*,partopsep=0pt,topsep=0pt]
    \item Through UAV measurements in densely-built areas, an urban 3D radio map dataset is developed, which allows us to evaluate the performance of 3D map estimations. Our real data can serve as field-test guidance for future works on 3D radio map estimation.
    \item To estimate urban 3D radio maps, based on Gaussian Process Regression (GPR), a joint model-and-data-driven scheme is proposed. Utilizing prior information provided by simulations, our scheme achieves higher estimation accuracy than the state-of-the-art schemes.
    \item Selecting proper measurement/training points is essential for map estimation. To this end, an online MAP-based method and an offline clustering-based method are proposed for training point selection, both of which can significantly reduce the measurement cost of 3D radio map recovery.
\end{itemize}

The rest of the paper is organized as follows. Section \ref{sec: 3D Radio Map Dataset} introduces a 3D radio map dataset collected by UAVs in densely-built urban environments. Section \ref{sec: Estimation Algorithm} proposes a GPR-based model-and-data-driven 3D radio map estimation scheme for sparse measurements and provide both offline and online methods to guide UAVs in selecting measurement points. In Section \ref{sec: Performance Evaluation}, results are provided for quantifying the performance of our proposed scheme. Finally, conclusions are drawn in Section \ref{sec: Conclusion}.

\vspace{-3mm}
\section{3D Radio Map Dataset}
\label{sec: 3D Radio Map Dataset}

In this section, we introduce a 3D radio map dataset for densely-built urban environments, built through the data collected by UAVs. To provide a comprehensive overview, we detail the measurement scenario selection, the UAVs, measurement devices, and the characteristics of the dataset as follows.
\vspace{-5mm}
\subsection{Measurement Scenario Selection}
\vspace{-1mm}

\begin{figure}[!t]
\vspace{-5mm}
	\center
	\includegraphics[width=0.85\linewidth, keepaspectratio]{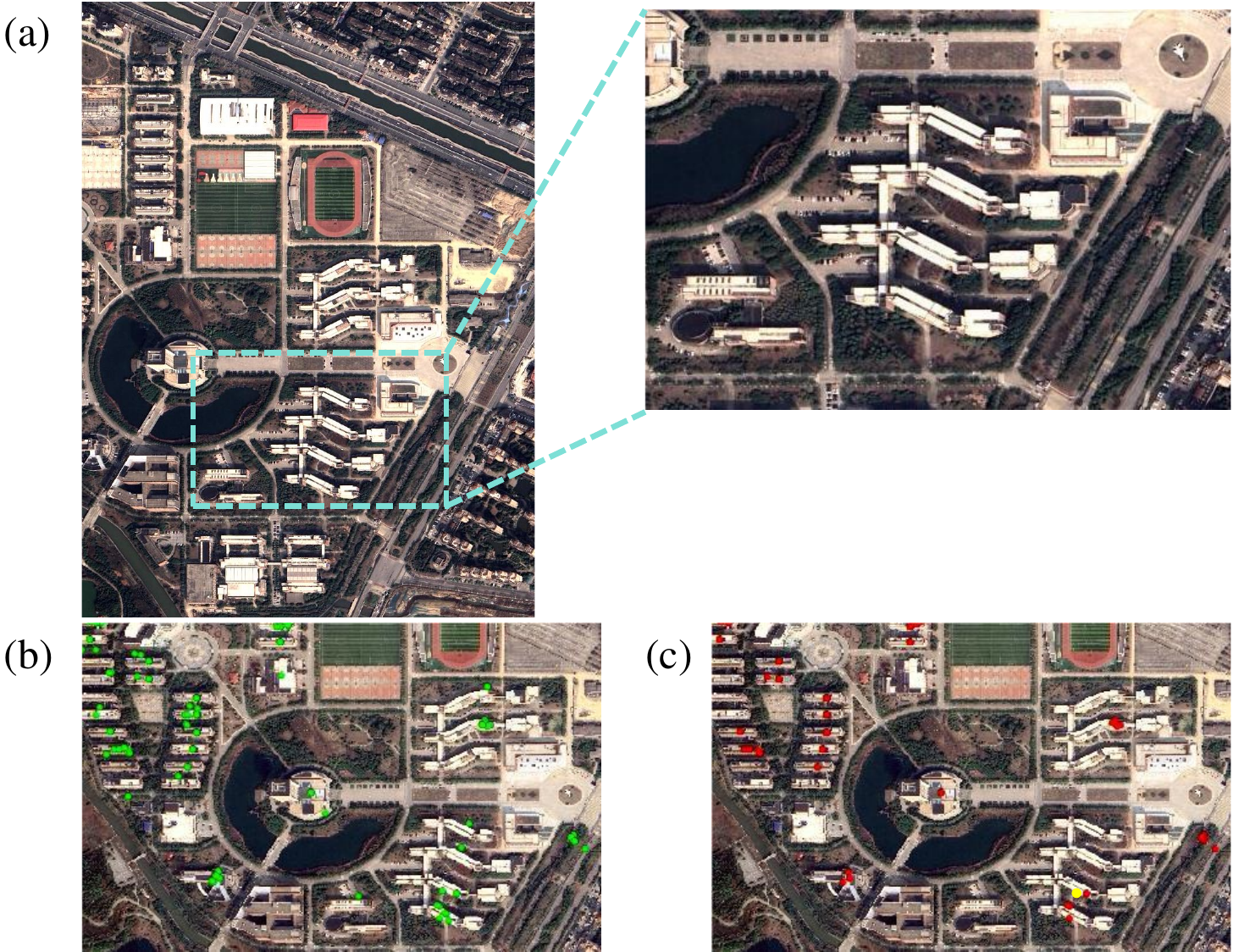}	
        \vspace{-2mm}
        \caption{Satellite images: (a) Overview of the campus; (b) Distribution of 5G base stations; (c) Distribution of 4G base stations.}
	\label{BplusS}

	\vspace{-2mm}
\end{figure}

Considering the application scenarios of UAVs in logistics field, which often involve densely-built areas, we select the teaching building area of a campus in Nanchang for radio map data collection due to its typical urban characteristics. As illustrated in Fig. \ref{BplusS}, the blue dashed rectangle on the left highlights the data collection region, with a magnified view on the right. This region provides a representative urban environment with various obstacles, both man-made (buildings, cars, walls) and natural (trees and vegetation), that affect radio signal propagation. Additionally, as shown in Fig. \ref{BplusS}(a), we obtain the latitude and longitude of base stations from the operator. Fig. \ref{BplusS}(b) displays the distribution of 5G base stations within the campus marked by green dots, while Fig. \ref{BplusS}(c) illustrates the distribution of 4G base stations marked by red dots.

\vspace{-5mm}
\subsection{UAV and Measurement Devices}
\vspace{-1mm}
\begin{figure}[!t]
\vspace{-1mm}
	\center
	\includegraphics[width=0.85\linewidth, keepaspectratio]{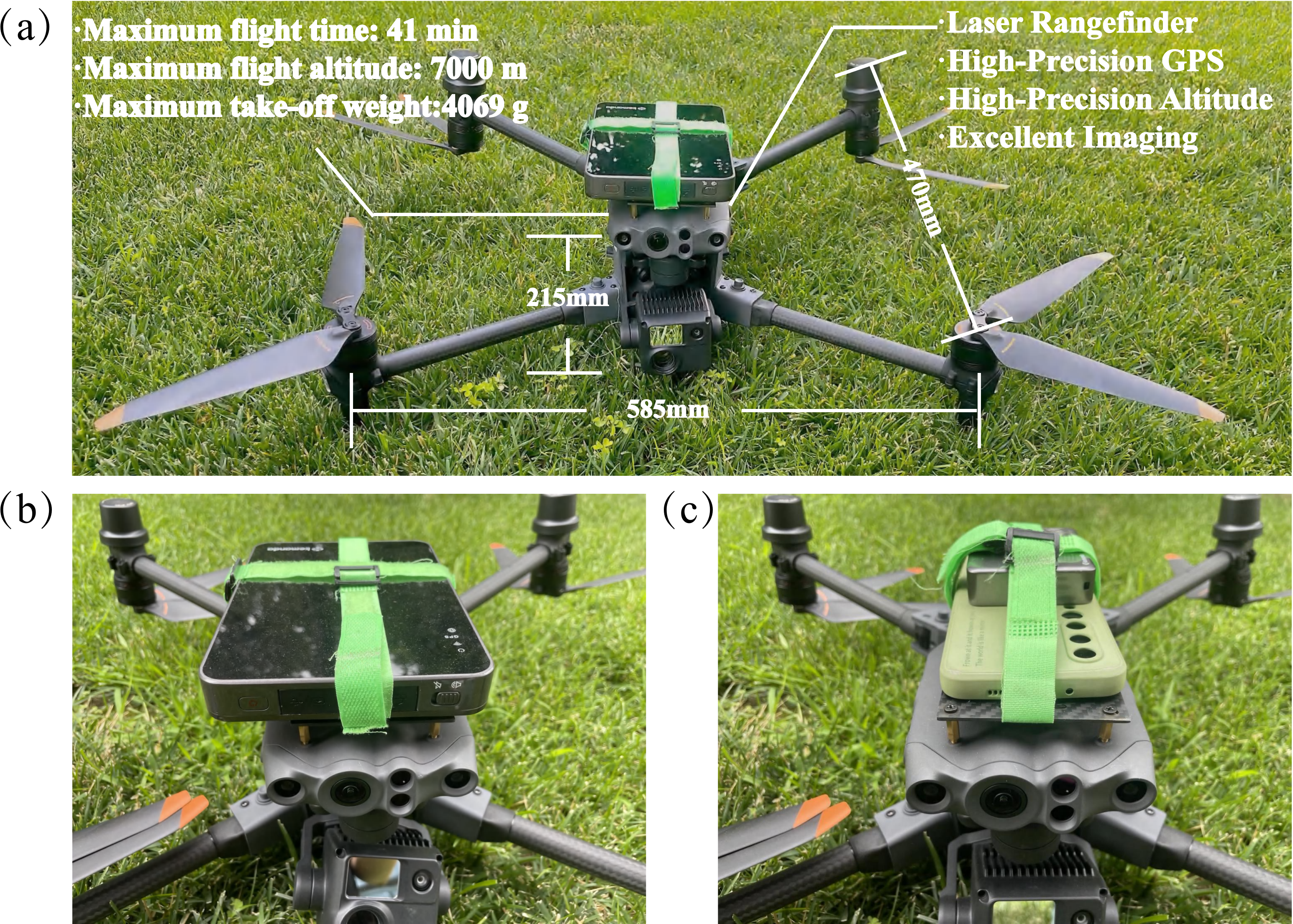}	
        \vspace{-2mm}
        \caption{UAV and measurement devices: (a) UAV equipped with measurement devices; (b) Base station switchable device; (c) Base station fixed device.}
	\label{UAV+Mea}
	\vspace{-5mm}
\end{figure}
{\it 1) UAV:} As illustrated in Fig. \ref{UAV+Mea}(a), the UAV, DJI Matrice 30 RTK, equipped with the measurement devices used in our experiments, is shown along with the relevant parameters.
{\it 2) Measurement Devices:} Fig. \ref{UAV+Mea}(b) and (c) illustrate the measurement devices used in our experiments. The black device in Fig. \ref{UAV+Mea}(b) can record the UAV's handovers between base stations, tracking connectivity changes as it navigates through the urban environment. The gray device in Fig. \ref{UAV+Mea}(c) is set to concentrate on a specific base station and can continuously record Reference Signal Received Power (RSRP).
\vspace{-5mm}
\subsection{Dataset Composition} 

\begin{figure}[!t]
\vspace{-3mm}
	\center
	\includegraphics[width=0.9\linewidth, keepaspectratio]{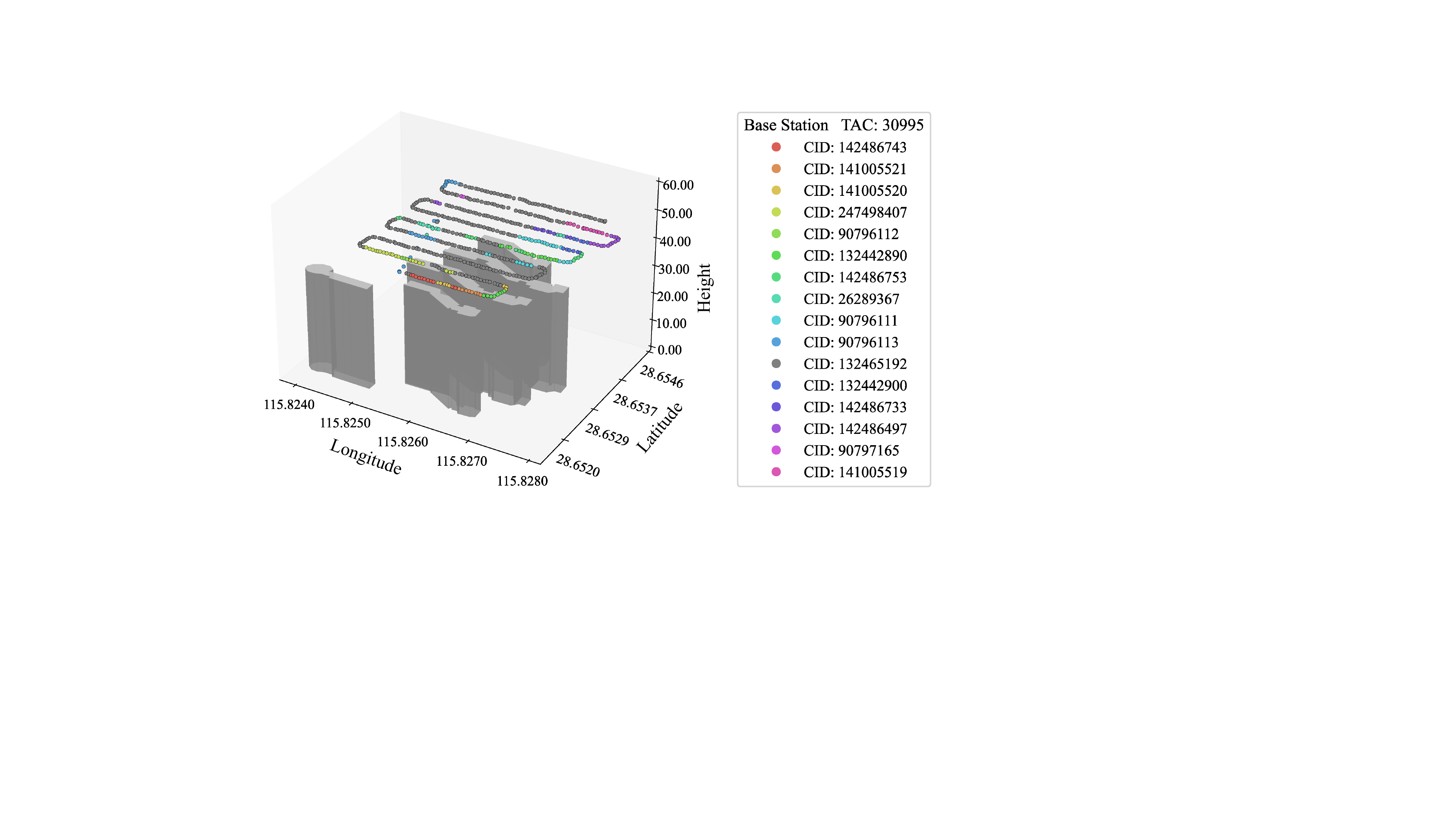}	
    \vspace{-3mm}
        \caption{Base station handovers while UAV's flying over the data collection region.}
	\label{switch}
	\vspace{-5mm}
\end{figure}

\begin{figure*}[!t]
	\center            \includegraphics[width=0.87\linewidth, keepaspectratio]{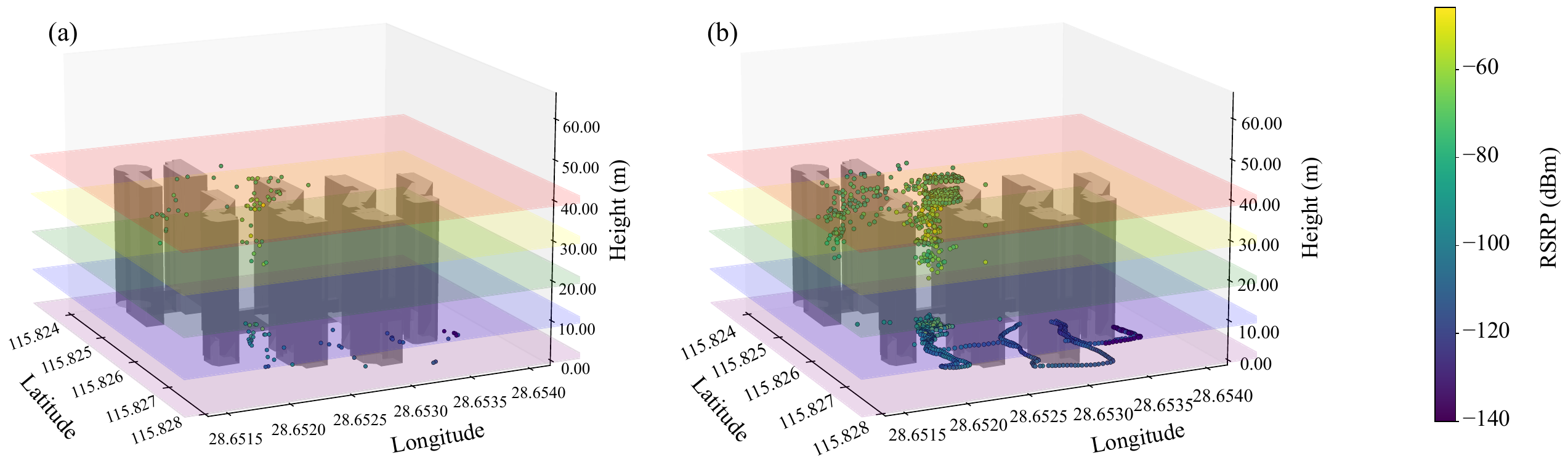}	
    \vspace{-3mm}
        \caption{3D radio maps obtained by UAV measurements: (a) Measurements for training (10\%); (b) Overall measurements (100\%).}
	\label{measurement}
	\vspace{-5mm}
\end{figure*}

In the data collection region, we use the switchable device to simulate a mobile phone, capturing base station handovers and confirming the set of receivable base stations. The results, shown in Fig. \ref{switch}, demonstrate frequent base station handovers during continuous UAV flights over the region. Using the base station distribution provided by the operator, we identify the positions of these receivable base stations and select the one that indicated by the yellow dot in Fig. \ref{BplusS}(c) as the transmitter for the radio map dataset. The UAV, equipped with the fixed device, then navigates the region, continuously collecting the RSRP of the base station. The spatial distribution of the RSRP in the 3D radio map dataset is shown in Fig. \ref{measurement}. The measurement device provides real-time latitude, longitude, and RSRP, while the UAV's log provides altitude information.

To enhance the dataset, we incorporate {\it simulated RSRP} for a comprehensive comparison with real-world measurements. Using Wireless Insite\footnote{\url{https://www.remcom.com/wireless-insite-propagation-software}}, we simulate RSRP based on 3D building models from Open Street Map\footnote{\url{https://www.openstreetmap.org}} in the data collection region. The simulation provides a theoretical model of signal propagation, accounting for the physical structures and environmental conditions of our measurements. The parameters for measurements and simulations are summarized in Table \ref{t_dataset}.
\renewcommand{\footnoterule}{
  \vspace{0.1cm} 
  \hrule width \linewidth 
  \vspace{0.1cm} 
}

\begin{table}[!t]
	\centering
        \vspace{-0mm}
	\caption{Parameters of the 3D Radio Map Dataset}
        \vspace{-2mm}
	\label{t_dataset}
	\begin{tabular}{cc}
		\hline
		\hline
		\textbf{{Parameters}} & \textbf{Value}\\
		\hline
            \multicolumn{2}{c}{Common Parameters for Measurement and Simulation} \\
            \hline
            Area (length $\times$ width) & 300 [m] $\times$ 280 [m] \\
            Height span & 0 [m] - 50 [m]\\
            Height range of the buildings & 6 [m] - 38.6 [m] \\ 
            Tx height & 2 [m] above the rooftop \\
            Center carrier frequency & 2.645 [GHz]\\
            Channel bandwidth & 20 [MHz] \\ 
            \hline
            \multicolumn{2}{c}{Measurement Parameters} \\
            \hline
            Data volume & 4274 \\
            \hline
            \multicolumn{2}{c}{Simulation Parameters} \\
            \hline
            Propagation model & Full 3-D \\
            Ray tracing method & Shooting and Bouncing Rays \\
            Waveform & Sinusoid \\ 
		\hline\hline
		
	\end{tabular}
\vspace{-0.4cm}
\end{table}

\vspace{-2mm}
\section{GPR-based 3D Radio Map Estimation Scheme}

\label{sec: Estimation Algorithm}

In this section, the GPR-based map estimation scheme is proposed to recover a full 3D radio map from a small number of UAV measurements, i.e., sparse data. 

\vspace{-5mm}
\subsection{Preliminary}
We consider the downlink transmission of a cellular network to focus on the coverage of a single transmitter. The RSRP reflects path loss laws within the solid angle visible to the user equipment. Since fast-fading effects have been averaged out over time by the receivers during measurements, environmental effects, such as diffraction, reflection, and absorption, are slow-fading effects that contribute to signal variations \cite{sohrabi2017construction}. For \(N\) positions \(\mathcal{P}^N := \{ \mathbf{p}_1, \dots, \mathbf{p}_N \}\), where \(\mathbf{p}_n := [p_x, p_y, p_z]^\mathrm{T} \in \mathbb{R}^3\), the RSRP at position \(\mathbf{p}_n\), denoted as \(\gamma(\mathbf{p}_n) \in \mathbb{R}\), can be modeled as
\vspace{-1mm}
\begin{equation}
    \begin{aligned}
        \gamma(\mathbf{p}_n) = \hat{\gamma}(\mathbf{p}_n) + r(\mathbf{p}_n) + z(\mathbf{p}_n),
    \end{aligned}
    \label{RSRP}
\end{equation}
where:
\(\hat{\gamma}(\mathbf{p}_n)\) represents the RSRP prediction from ray-tracing (simulated RSRP, in dB), 
\(r(\mathbf{p}_n)\) is the residual, which accounts for the prediction errors due to factors not captured by ray-tracing, such as slow fading caused by unmodeled environmental effects,
\(z(\mathbf{p}_n)\) represents the additive zero-mean noise at position \(\mathbf{p}_n\).

Our goal is to estimate the RSRP of the 3D radio map for a base station. For the ease of expression, we provide the following definitions:
\begin{itemize}[leftmargin=*,partopsep=0pt,topsep=0pt]
    \item \textbf{Base Station Information}: The location, center carrier frequency and channel bandwidth of the base station.
    \item \textbf{Simulated RSRP}: The dataset comprises $M$ training points $\mathcal{P}^M := \{ \mathbf{p}_1, \dots, \mathbf{p}_M \}$ and $N-M$ testing points $\mathcal{\bar{P}}^{N-M} := \{ \mathbf{p}_{M+1}, \dots, \mathbf{p}_N \}$. The simulated RSRP at position $\mathbf{p}_n$ is denoted as  $\hat{\gamma}(\mathbf{p}_n)$, where $n \in \{1, \dots, N\}$ and $N\gg M$.  
    \item \textbf{Measured RSRP}: The RSRP $\gamma(\mathbf{p}_m)$ measured at the $M$ training points $\mathcal{P}^M$, where $m \in \{1, \dots, M\}$.

\end{itemize}

For example, Fig. \ref{measurement}(a) shows a map of RSRP measurements, where 10\% of the measurements are available for training. Fig. \ref{measurement}(b) displays the actual 3D RSRP map obtained from UAV measurements. RSRP, represented by \(\hat{\gamma}(\mathbf{p}_n)\) in (\ref{RSRP}), is derived from ray-tracing using Wireless InSite, with unknown transmission power. Note that, although ray tracing accurately captures path loss, it cannot fully simulate the nuances of real-world slow fading, leading to inaccuracies in \(\hat{\gamma}(\mathbf{p}_n)\). Our aim is to use \(M\) RSRP measurements in Fig. \ref{measurement}(a) and \(N-M\) simulated RSRP values to recover the overall RSRP map in Fig. \ref{measurement}(b).

Thus, our problem becomes estimating the slow-fading components of a 3D RSRP map. They are influenced by environmental details that ray-tracing methods cannot capture and should be determined based on sparse measurement data. Slow-fading effects depend on the local environment's topological structure, such as land use type, nearby obstructions like trees and buildings, and line of sight conditions. Determined by environmental effects, there is a spatial correlation between the RSRP at close-by locations. Therefore, modeling the residual $r(\mathbf{p})$ as a stochastic process offers an elegant means of specifying function properties. The residual $r(\mathbf{p})$ is defined as the difference between the measured RSRP $\gamma(\mathbf{p})$ and the simulated RSRP $\hat{\gamma}(\mathbf{p})$, given by
\begin{equation}
    \begin{aligned}
        r(\mathbf{p}) = \gamma(\mathbf{p}) - \hat{\gamma}(\mathbf{p}).
    \end{aligned}
    \label{residual}
   \vspace{-2mm}
\end{equation}
\subsection{GPR-based 3D Radio Map Estimation Scheme}
A Gaussian process (GP), denoted as $\mathcal{GP}(\mu(\mathbf{x}), k(\mathbf{x}, \mathbf{x'}))$, is a collection of random variables where any finite subset follows a multivariate Gaussian distribution. It is fully determined by its mean function $\mu(\mathbf{x})$, often assumed to be zero, and its covariance kernel $k(\mathbf{x}, \mathbf{x'})$, which dictates the smoothness characteristics of the function it models. In our proposed GPR-based scheme, the residual $r(\mathbf{x})$ is modeled as a GP. Since the close-by locations may exhibit significant differences in signal strength due to diverse building shapes, the input $\mathbf{x}$ includes both the locations and the simulated RSRP, i.e., $\mathbf{x} = [p_x, p_y, p_z, \hat{\gamma}(p_x, p_y, p_z)]^\mathrm{T}$. This dimension extension ensures that spatially adjacent points with similar environmental characteristics have high correlations, enabling localized GPR to effectively handle non-stationary data.

Consider a prior $\mathcal{GP}(\mu(\mathbf{x}), k(\mathbf{x}, \mathbf{x'}))$ over $r(\mathbf{x})$. Let $y(\mathbf{x})$ denote the RSRP measurement at points in \(\mathcal{X}^M := \{ \mathbf{x}_1, \dots, \mathbf{x}_M \}\), and $\hat{y}(\mathbf{x})$ denote the simulated RSRP at those points. Thus, the residual $r(\mathbf{x})$ is given by
\begin{equation}
    r(\mathbf{x}) = y(\mathbf{x}) - \hat{y}(\mathbf{x}).
\end{equation}
The joint probability distribution of the residuals at measured (training) points $\mathcal{X}^M := \{ \mathbf{x}_1, \dots, \mathbf{x}_M \}$ and unmeasured (testing) points $\mathcal{X}^{N-M} := \{ \mathbf{x}_{M+1}, \dots, \mathbf{x}_N \}$ is
\begin{equation}
    \begin{bmatrix}
        r(\mathbf{x}_*) \\
        \boldsymbol{r}
    \end{bmatrix}
    \sim \mathcal{N} \left(
    \begin{bmatrix}
        \mu(\mathbf{x}_*) \\
        \boldsymbol{\mu}
    \end{bmatrix},
    \begin{bmatrix}
        \Sigma(\mathbf{x}_*) & \mathbf{k}(\mathbf{x}_*)^\mathrm{T} \\
        \mathbf{k}(\mathbf{x}_*) & \mathbf{K}
    \end{bmatrix}
    \right),
    \label{joint_distribution}
\end{equation}
where $\mathbf{x}_* \!\in\! \mathcal{X}^{N-M}$, ${\mu}(\mathbf{x}_*)$ denotes the priori mean of the residual at $\mathbf{x}_*$, ${\bf{r}} := [ {r\left( {{{\bf{x}}_1}} \right), \cdots ,r\left( {{{\bf{x}}_M}} \right)} ]^{\rm T}$, $\mathbf{k}(\mathbf{x}_*) := [k(\mathbf{x}_1, \mathbf{x}_*), \cdots, k(\mathbf{x}_M, \mathbf{x}_*)]^\mathrm{T}$, $\boldsymbol{\mu}:= [\mu(\mathbf{x}_1), \cdots, \mu(\mathbf{x}_M)]^\mathrm{T}$, and the $(i,j)$-th entry of $\mathbf{K} \in \mathbb{C}^{M \times M}$ is $k(\mathbf{x}_i, \mathbf{x}_j)$. 

Thus, the posterior over $r(\mathbf{x}_*)$ is also a Gaussian process, with mean and covariance can be derived from (\ref{joint_distribution}) as
\begin{equation}
    {\mu}_*(\mathbf{x}_*) = {\mu}(\mathbf{x}_*) + \mathbf{k}(\mathbf{x}_*)^\mathrm{T} \mathbf{K}^{-1} ({\bf r} - \boldsymbol{\mu}),
    \label{posterior_mean}
\end{equation}
\begin{equation}
    \vspace{-2mm}
    {\Sigma}(\mathbf{x}_*) = k(\mathbf{x}_*, \mathbf{x}_*) - \mathbf{k}(\mathbf{x}_*)^\mathrm{T} \mathbf{K}^{-1} \mathbf{k}(\mathbf{x}_*).
    \label{posterior_covariance}
\end{equation}
With sufficient training points, ${\mu}_*(\mathbf{x}_*)$ can be viewed as the Bayesian estimator of $r(\mathbf{x}_*)$, which can achieve our goal.

In GPR, the wide-sense stationarity of data significantly impacts model performance. Non-stationary data require a covariance function that accounts for varying statistical properties, which increases model complexity and poses challenges for parameter estimation. These factors hinder GPR's generalization and prediction accuracy. To mitigate these issues, we employ a composite kernel function. This function combines the Constant Kernel ($k_{\text{const}}$), Matérn Kernel ($k_{\text{Matérn}}$), and White Noise Kernel ($k_{\text{WN}}$), and is expressed as follows:
\begin{equation}
    k_{\text{com}}(\mathbf{x}, \mathbf{x'}) \!=\! k_{\text{const}}(\mathbf{x}, \mathbf{x'}) \!\times\! k_{\text{Matérn}}(\mathbf{x}, \mathbf{x'})\!+\!k_{\text{WN}}(\mathbf{x}, \mathbf{x'}),
\end{equation}
as detailed in \cite[Eq. (4.14)]{Rasmussen2006Gaussian}.

Selecting proper measurement/training points is crucial for 3D radio map estimation as it directly impacts the accuracy, representativeness, resource efficiency, and robustness of the model. Below we introduce an online and an offline method for selecting measurement points. The online method continuously updates strategies based on real-time results, with subsequent decisions depending on feedback data. In contrast, the offline method does not require real-time feedback during execution and relies solely on pre-existing knowledge and data. 

\subsubsection{Online Point Selection Method}
In GPR, the posterior variance at each point quantifies the uncertainty of the model's predictions. High posterior variance indicates greater uncertainty and more potential information. Thus, selecting the next measurement point associated with the MAP variance can maximize the information acquisition. To this end, the candidate point ${{\bf x}_{t+1}}$ can be chosen according to
\begin{equation}
    \begin{aligned}
   \vspace{-2mm}
    {{\bf x}_{t+1}} = \arg\!\!\!\!\!\!\max_{\mathbf{x} \in \mathcal{X}^N / \mathcal{X}^t} \Sigma^t(\mathbf{x}),
    \end{aligned}
    \label{t+1}
   \vspace{-2mm}
\end{equation}
where $\mathcal{X}^N$ represents all potential measurement points; and $\mathcal{X}^t$ represents measured points; and $\Sigma^t(\mathbf{x})$ is the posterior covariance derived from the measured points $\mathcal{X}^t$ by (\ref{posterior_covariance}). The online point selection algorithm is detailed in {\bf Algorithm~\ref{alg: online}}.

\begin{algorithm}[!t] 
\small
\caption{Online Point Selection Using GPR with MAP}
\label{alg: online}
\begin{algorithmic}[1]
\REQUIRE The candidate set $\mathcal{X}^N := \{ \mathbf{x}_n \}_{n=1}^N$, the target number $M$ for the training set.
\STATE Randomly select $\mathbf{x}_1$ from $\mathcal{X}^N$ and initialize $\mathcal{X}_{\text{train}} := \{ \mathbf{x}_1 \}$.
\STATE Initialize the remaining candidate set $\mathcal{X}_{\text{cand}} := \mathcal{X}^N / \mathcal{X}_{\text{train}}$.
\WHILE{$t\in\{1,\cdots,M-1\}$}
    \STATE Using the training set $\mathcal{X}_{\text{train}}$, calculate the posterior covariance $\Sigma^t(\mathbf{x})$ for all $\mathbf{x} \in \mathcal{X}_{\text{cand}}$ via (\ref{posterior_covariance}).
    \STATE Select the candidate point:
    $
    \mathbf{x}_{t+1} = \arg\!\max_{\mathbf{x} \in \mathcal{X}_{\text{cand}} } \Sigma^t(\mathbf{x})
    $.
    \STATE Update the training set: $\mathcal{X}_{\text{train}} \leftarrow \mathcal{X}_{\text{train}} \cup \{ \mathbf{x}^{t+1}\}$.
    \STATE Update the candidate set: $\mathcal{X}_{\text{cand}} \leftarrow \mathcal{X}_{\text{cand}} / \{\mathbf{x}^{t+1}\}$.
\ENDWHILE
\STATE Obtain training set $\mathcal{X}^M:=\mathcal{X}_{\text{train}}$.
\ENSURE The designed training set $\mathcal{X}^M := \{ \mathbf{x}_i \}_{i=1}^M$.
\end{algorithmic}
\end{algorithm}
\begin{algorithm}[!t] 

\small
\begin{algorithmic}[1]
\caption{Offline Point Selection Using KMeans Clustering}
\label{alg: offline}
\REQUIRE The candidate set $\mathcal{X}^N := \{ \mathbf{x}_n \}_{n=1}^N$, the target number $M$ for the training set.
\STATE Initialize KMeans with $M$ clusters and fit it to candidate set $\mathcal{X}^N$.
\STATE Assign all candidate points to $M$ clusters via KMeans.
\FOR{each cluster $l \in \{1, 2, \dots, M\}$}
    \STATE Get the centroid of the $l$-th clusters by ${{\bf{c}}_l} = ( {\sum\nolimits_{{\bf{x}} \in {{\cal X}_l}} {\bf{x}} } )/\left| {{{\cal X}_l}} \right|$ where $\mathcal{X}_l$ is the set of candidate points in cluster $l$.
    \STATE Find the candidate point:
    ${{\bf{x}}_{{n_l}}} = \mathop {\arg\!\min }\nolimits_{{\bf{x}} \in {\cal X}_l} \left\| {{\bf{x}} - {{\bf{c}}_l}} \right\|$.
\ENDFOR
\STATE Obtain training set $\mathcal{X}^M = \{ \mathbf{x}_{n_1}, \mathbf{x}_{n_2}, \dots, \mathbf{x}_{n_M} \}$.
\ENSURE The designed training set $\mathcal{X}^M := \{ \mathbf{x}_i \}_{i=1}^M$.
\end{algorithmic}
\end{algorithm}

\subsubsection{Offline Point Selection Method}
Although the MAP method is effective, it is constrained by its requirement for real-time computation and its exclusive focus on GPR. Given the spatial correlation of environmental factors in 3D radio maps, we propose an offline KMeans-based method to select a series of measurement points simultaneously. By clustering the points by \(\mathbf{x} = [p_x, p_y, p_z, \hat{\gamma}]\), and employing cluster centroids as measurement points, we maximize sampling diversity and enhance the efficiency and applicability of 3D radio map estimation. The specific algorithm is detailed in {\bf Algorithm~\ref{alg: offline}}.

\vspace{-2mm}
\section{Performance Evaluation}
\label{sec: Performance Evaluation}

\begin{table}[!t]
\centering
\vspace{-2mm}
\caption{Comparison of Kernel Function Performance}
\vspace{-2mm}
\label{Kernel}
\begin{tabular}{|c|c|c|} 
\hline
\textbf{Kernel Type} & \textbf{Kernel Selection} & \textbf{RMSE (dB)} \\ 
\hline
\multirow{4}{*}{}         & $k_\text{RBF}$ + $k_\text{WN}$  & 23.9106   \\
                    Radial Basis      & $k_\text{RBF}$  & 17.5451                                        \\
                    Function      & $k_\text{const}$ $\times$ $k_\text{RBF}$  & 14.4392                                        \\
                          & $k_\text{const}$ $\times$ $k_\text{RBF}$ + $k_\text{WN}$  & 8.2593                                         \\ 
\hline
\multirow{4}{*}{}         & $k_\text{RQ}$ + $k_\text{WN}$  & 23.9106                \\
                    Rational      & $k_\text{RQ}$  & 10.3063                            \\
                    Quadratic      & $k_\text{const}$ $\times$ $k_\text{RQ}$ & 9.1089                                         \\
                          & $k_\text{const}$ $\times$ $k_\text{RQ}$ + $k_\text{WN}$ & 8.0131                                         \\ 
\hline
\multirow{4}{*}{}         & $k_\text{Matérn}$ + $k_\text{WN}$  & 23.9107                                        \\
                          & $k_\text{Matérn}$   & 8.8130                                         \\
                    Matérn      & $k_\text{const}$ $\times$ $k_\text{Matérn}$ & 8.0265                                         \\
                          & {\cellcolor[rgb]{0.8,0.8,0.8}}\textbf{$k_{\text{const}}$ $\times$ $k_{\text{Matérn}}$ + $k_{\text{WN}}$} & {\cellcolor[rgb]{0.8,0.8,0.8}}\textbf{7.9286}  \\
\hline
\end{tabular}
\vspace{-0.4cm}
\end{table}

In this section,  we evaluate the performance of the proposed GPR-based scheme on our built dataset. 

We first evaluate the effect of various kernel combinations on modeling environmental factors and slow-fading effects influenced by local topological structures. Radial Basis Function ($k_{\text{RBF}}$), Matérn ($k_{\text{Matérn}}$), and Rational Quadratic ($k_{\text{RQ}}$) kernels are used to address smoothness, roughness, and multiple variation scales. Additionally, Constant ($k_{\text{const}}$) and White Noise ($k_{\text{WN}}$) kernels are included to account for constant bias and inherent measurement noise, enhancing the robustness and accuracy of the GPR model. As shown in Table \ref{Kernel}, the best performance is achieved with a combination of $k_{\text{const}}$, $k_{\text{Matérn}}$, and $k_{\text{WN}}$ due to the Matérn kernel's tunable smoothness parameter. Thus, in the following simulations, we use this setup to enable our proposed scheme.

The following three baselines are considered:

\begin{itemize}[leftmargin=*,partopsep=0pt,topsep=0pt]

    \item \textbf{Inverse Distance Weighting (IDW)}: A local neighborhood method that interpolates spatial points based on their influence within a certain distance \cite{denkovski2012reliability}.

    \item \textbf{K-Nearest Neighbors (KNN)}: A classical machine learning algorithm for intensity prediction. It interpolates unknown values by identifying the \( k \) nearest training samples \cite{sohrabi2017construction}.
    \item \textbf{Kriging}: A geostatistical method that interpolates unknown signal power by considering distances and modeling spatial correlation with a variogram\cite{ivanov2023limited}.
\end{itemize}

\begin{figure}[!t]
	\center
	\includegraphics[width=0.96\linewidth, keepaspectratio]{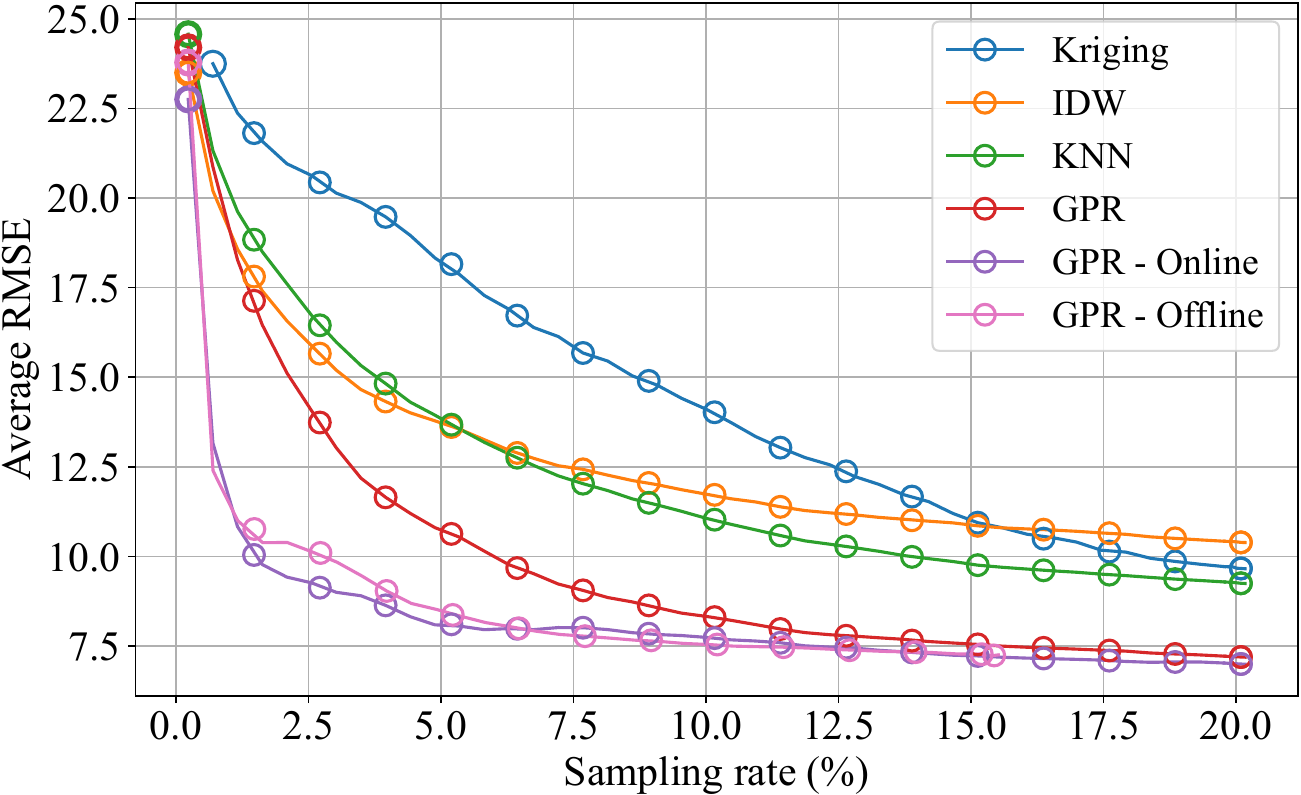}	
    \vspace{-2mm}
        \caption{Comparison of average RMSE as a function of sampling rate for different radio map estimation schemes. GPR schemes based on random point selection, online method, and offline method are compared.}
	\label{max}
	\vspace{-5mm}
\end{figure}

Fig. \ref{max} illustrates the average RMSE for different estimation schemes. All estimation schemes employ random point selection with simulated RSRP included for comparison, whereas GPR additionally employs the proposed online and offline point selection method. Three different point selection curves for GPR are presented: random (red), online MAP-based GPR (purple), and offline clustering-based GPR (pink). One can find that, our GPR-based scheme achieves the lowest RMSE across all sampling rates. It also outperforms other state-of-the-art schemes even after the plateau of curves, with an RMSE that is 2.5 dB lower than other schemes. Remarkably, both the offline and online methods achieve an RMSE of 10 dB with only 2\% of the sampling points, while other schemes require over 14\% of the sampling points to achieve the same performance. Additionally, the offline method shows performance comparable to the best-performing online method, significantly surpassing the random selection. This phenomenon highlights that the clustering-based method enables batch point selection using pre-existing knowledge and data, incurring minimal performance loss compared to the MAP-based method.

\begin{figure}[!t]
    \centering
    \includegraphics[width=0.96\linewidth, keepaspectratio]{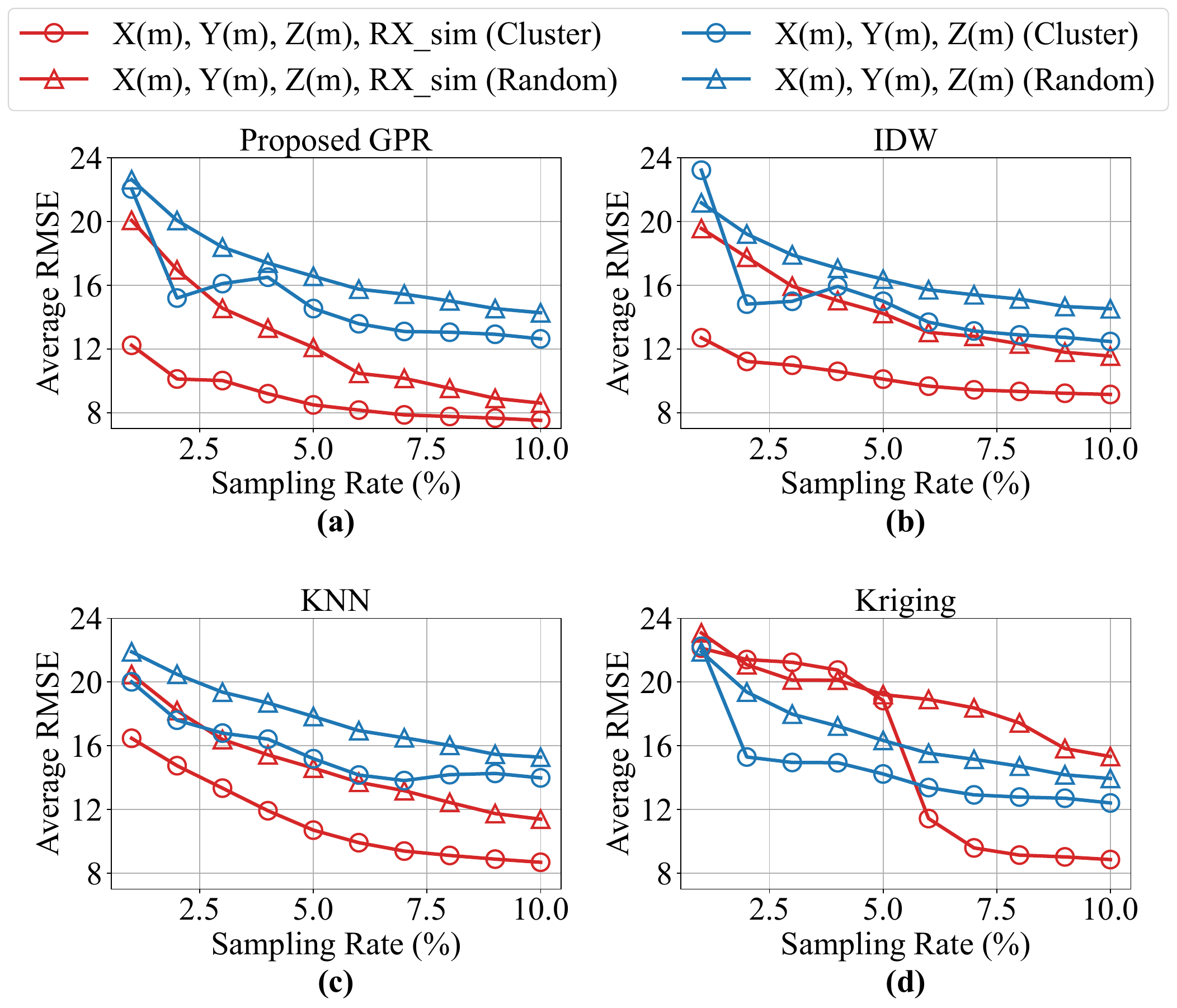}
    \vspace{-2mm}
    \caption{Comparisons of average RMSE as a function of sampling rates using different estimation schemes, input features, and point selection methods.}
    \label{4M}
   \vspace{-5mm}
\end{figure}

We conduct additional experiments to show the generality of our two proposed designs: The incorporation of simulated RSRP alongside spatial coordinates, and the use of clustering-based point selection, across various schemes. As shown in Fig. \ref{4M}, the inclusion of simulated RSRP consistently enhances estimation performance across all four schemes, with RMSE reductions of up to 10 dB. The clustering-based method shows a marked improvement in performance, which reduces RMSE by up to 8 dB compared to the random selection method. These findings underscore the generality of these two designs in integrating with other 3D radio map estimation schemes.

\vspace{-1mm}
\section{Conclusion}
\label{sec: Conclusion}
In this paper, we develop a 3D UAV-collected dataset in urban scenarios and propose a GPR-based platform for 3D radio map estimation. Posterior variance calculation is incorporated into the GPR framework for an online MAP point selection method.
This integration enables more efficient and accurate estimations with sparse measurements by prioritizing points with higher uncertainty, resulting in the best performance observed. Additionally, we propose an offline clustering-based point selection method to ensure representative training points, which performs as well as its online counterparts. This efficient platform is crucial for logistics UAVs, as accurate 3D radio maps have become essential for reliable navigation and communication in densely-built urban environments. We also prove the generality of our designs, thereby providing a framework for future works on 3D radio map estimation.

\ifCLASSOPTIONcaptionsoff
  \newpage
\fi

\bibliographystyle{IEEEtran}
\bibliography{reference}

\end{document}